\documentclass[12pt]{article}
\usepackage{amssymb}
\usepackage{graphicx}
\usepackage{epstopdf}
\usepackage[applemac]{inputenc}
\usepackage[normalem]{ulem}

\begin{document}
\title{Electrophoretic separation of large DNAs using steric confinement}
\author{Jérôme Math\'e$^1$, Jean-Marc Di Meglio$^{*,2}$ and Bernard Tinland$^3$}
\maketitle

\normalsize

$^1$Mat\'eriaux et Polym\`eres aux Interfaces, Universit\'e d'\'Evry-Val d'Essonne, 91025 \'Evry Cedex,France\\

$^2$Mati\`ere et Syst\`emes Complexes, UMR 7057 CNRS \& Universit\'e Paris Diderot, 75205 Paris Cedex 13, France\\

$^3$Centre de Recherche en Mati\`ere Condens\'ee et Nanosciences - UPR 7251 CNRS - 13288 Marseille Cedex 09 - France\\
\vskip 3cm
\small {*Corresponding author :\\
email address: jean-marc.dimeglio@univ-paris-diderot.fr\\
Postal address : MSC, Université Paris Diderot, Bâtiment Condorcet, CC7056,\\
10 rue Alice Domon et Léonie Duquet, 75205 Paris Cedex 13\\
Telephone: (33) 57 27 61 97 Fax: (33) 1 57 27 62 11}

\newpage
\normalsize
\section*{Abstract}
We report an alternative method for electrophoretic separation of large DNAs using steric confinement
between solid walls, without gel or obstacles. The change of electrophoretic mobility {\em vs} confinement
thickness is investigated using fluorescence video microscopy. We observe separation at small confinement thicknesses
followed by a transition to the bulk behaviour (no separation) at a thickness of about 4 $\mu$m (a few radii of gyration for the studied DNA chains). We present tentative explanations of our original observations.\\

Keywords:\\
Electrophoresis, electro-osmosis, DNA

\newpage
\section{Introduction} Capillary electrophoresis in polymer solutions has become an alternative method of gel
electro\-pho\-resis for DNA separation (review in \cite{Slater}). Simultaneously, there has been a stimulating
development of new methods using new geometries. For instance, separation of DNA adsorbed on a flat surface
\cite{Raphail} exploits the dependence of the adsorption-desorption kinetics on chain length; microlithography
has been used to create topological defects and entropic trapping \cite{Craig}. Is it possible to find a
simpler way of separation? Could the sole geometrical confinement be sufficient? Two groups have worked in
this direction. Roeraade {\em et al.} \cite{Roeraade} have separated a mixture of HindIII-digested $\lambda$-DNA
fragments, using capillaries with diameters of order of two radii of gyration ($R_{g}$) of the largest
fragment. Iki {\em et al.} \cite{Iki} have used $20$~$\mu$m-diameter capillaries. Both groups have explained the separation
using electrostatics. Roeraade has considered that the smallest molecules are repelled towards the capillary center
since walls and DNA are both negatively charged. He has calculated that capillaries with small inner diameter ($<1$~$\mu$m) filled with low ionic strength solution lead to an electro-osmotic flow (EOF) with a Poiseuille
profile: the smallest molecules experience the fastest stream lines and are eluted first. Iki has considered that
the smallest molecules have a smaller mobility in the vicinity of the wall because of the screening by the
surface counter-ions; with the EOF in the opposite direction of the electrophoretic force, the smaller the
molecule, the higher velocity in the capillary.

The present paper reports experiments performed in different conditions (higher ionic strength than
\cite{Roeraade}, tighter confinement and much lower electric fields) that exhibit totally different behaviours. 

\section{Materials and methods}

\begin{figure}[ht]
\includegraphics[width=10cm]{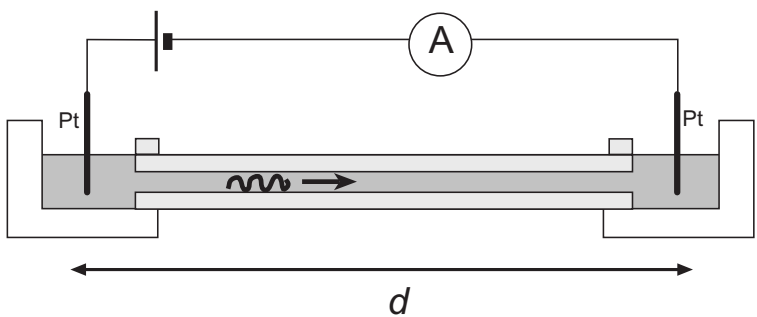}
\caption{Schematic representation of the measurement cell (the distance $d$ between electrodes is 8 cm)}
\label{f.1}
\end{figure}

Confinement is achieved using a home-built cell (Fig. \ref{f.1}). Two glass plates (microscopy slides) are
separated by the DNA solution and connected to two reservoirs, hosting Pt electrodes to apply an electric
field (using an ISO-TECH power supply). The glass plates are washed with 1 \% RBS solution (an alkaline powerful industrial detergent from Roth-Sochiel) in
a sonication bath (Branson 1200) for $10$~min, and then thoroughly rinsed with ultra pure water (Millipore
Milli-Q). The confinement gap is filled with a solution $36$ pmol$\cdot$l$^{-1}$  for $\lambda$-DNA (48.5 kbp
from Biolabs) or $11$ pmol$\cdot$l$^{-1}$ for $T2$-DNA (164 kbp from Sigma) in $10^{-2}$ mol$\cdot$ l$^{-1}$
TBE buffer solution $pH\,8.3$. Both concentrations correspond to the same mass concentration, $1.2$ $\mu$g$\cdot$ml$^{-1}$ (dilute regime). The DNAs' radii of gyration are estimated from the Benoit-Doty relation
\cite{Benoit}: $R_g \simeq \sqrt{L_c l_p/3}$ (with $L_c$ the contour length and $l_p$ the persistence length \cite{hager}). 
This gives $R_g = 0.56$~$\mu$m for $\lambda$-DNA ($L_c = 16.5$~$\mu$m and $l_p= 50$~nm) and
$R_g=1.13$~$\mu$m for $T2$-DNA ($L_c= 55.8$~$\mu$m and $l_p  = 50$~nm). The DNA molecules have been
labeled with a fluorescent dye (YOYO-1, Molecular Probes):  1 dye per 50 base-pairs. This very small dye
content ensures that the characteristics of DNA ({\em} charge, persistence length etc.) remains unaltered.
The images are taken at 25 frames per second during 8 seconds using a silicon intensified target (SIT) camera (Hamamatsu) fitted to a microscope (Leica). No
electric field is applied during the first two seconds to estimate the possible convection in the sample. Five
field values (from $\simeq$ 2 to 8~ V$\cdot$cm$^{-1}$) have been used and the electrophoretic mobility is
derived from the slope of the linear variation of velocity with applied field.

The image analysis is performed as follows. In a first set of experiments, using solutions containing only one
individual species (either $\lambda$-DNA or $T2$-DNA), we have derived the velocities of the chains from the
correlation of successive frames. In a second set of experiments we have investigated mixtures of both DNAs:
we have determined the position \cite{Grier} {\em and} the size of the molecules \cite{Maier} to derive the 2D
histogram of velocity {\em vs} radius and the mean velocity of each population.

The confinement thickness is determined from the electrical resistance of the confined liquid film. The
measurement of the current gives a constant and reproducible value 30 min after tightening the plates,
indicating mechanical equilibrium. The calibration is done using spacers of 0.5 and $2$~$\mu$m
(monodisperse polystyrene beads) and also gives the conductivity $C$ of the solution. Indeed the resistance
$R$ of the film is related to the thickness $e$ by the relation $R=\frac{{\textstyle 1}}{{\textstyle
C}}\frac{{\textstyle L}}{{\textstyle el}}$ where $L$ is the length of the plate along the direction of the
electric field, $l$ is the width of the plate normal to this field. We obtain a conductivity of $1.9 \pm
0.2$~mS$\cdot$cm$^{-1}$ in fairly good agreement with the $2.2$~mS$\cdot$cm$^{-1}$ value measured with a
conductimeter (Fisher scientific Accumet AR20) at room temperature. We set the gap thickness by tightening
uniformly the two plates together. The accuracy of the thickness measurement is around 10 \% and the thickness
is uniform within $\pm 0.1$~$\mu$m. The excess of solution leaving the gap on the side of the plates during
tightening is carefully removed to avoid any shortcut for the electric current. The evaporation of the solvent
is negligible because of the area of solvent exposed to air is very small. Finally, the variations of the
thickness explored by DNA molecules during an experiment of $8$~s are also considered as negligible.

\section{Results}

\begin{figure}[ht]
\includegraphics[width=10cm]{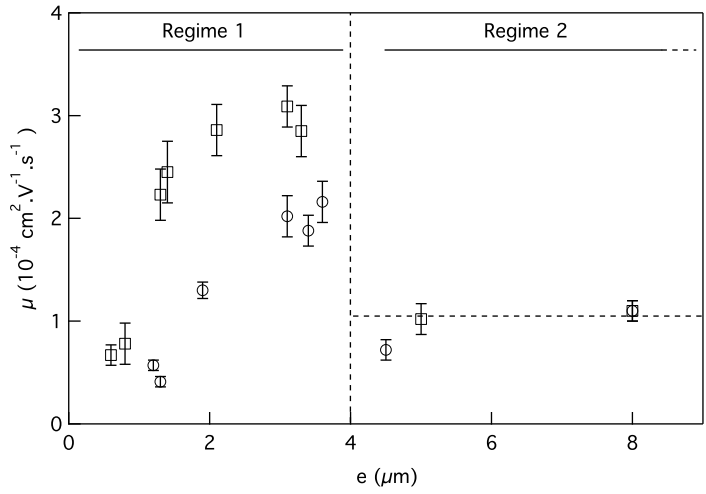}
\caption{Mobilities $\mu$ of $\lambda$ ({\large $\circ$}) and $T2$ ($\square$) DNA {\em vs} confinement thickness
$e$. The vertical dashed line represents the transition between the two regimes. The horizontal dashed line is a eye guide of the constant mobility in the second regime.}
\label{f.2}
\end{figure}

Fig. \ref{f.2} represents the measured mobility $\mu$ as a function of the confinement thickness $e$ (first
set of experiments). A first and striking feature is that $\mu$-values are positive: negatively charged DNAs move
towards the cathode in the same direction as the electric field $\mathbf{E}$. This behaviour is caused by the
presence of an electro-osmotic flow (EOF): glass plates are negatively charged at $pH=8.3$ and their
positively charged counter-ions, dragged by the electric field $\mathbf{E}$, induces a plug flow of the solution
with velocity  ${v_{EOF}}$ towards the negative electrode. This can be characterized by an electro-osmotic
mobility $\mu_{EOF}>0$ defined by ${v_{EOF}}=\mu_{EOF} {E}$. When this mobility is larger than the
intrinsic (and negative) electrophoretic mobility of DNA, the effective mobility is positive. 

We distinguish two regimes for $\mu(e)$. In a first regime, for \emph{small} thicknesses ($e < 4$~$\mu$m),
$\mu$ increases with a large dependence on chain length. Although $T2$-DNA is longer, it has a larger
mobility than $\lambda$-DNA: for instance, for $e=2$~$\mu$m, $T2$-DNAs move more than twice faster than
$\lambda$-DNAs.  EOF reverses the order of elution; this effect has
been already observed by Barron \cite{Barron} during DNA electrophoresis in very dilute polymer solutions
using uncoated capillary glass tubes. 

In a second regime, for thicknesses larger than $4$~$\mu$m, $\mu$ is constant and equal to  $+ (1\pm 0.3)
10^{-4}$~cm$^{2}\cdot$ V$^{-1}\cdot$s$^{-1}$ and neither depends on chain length nor confinement thickness.

The first and second regime are separated by a transition region with an important reduction of the mobility.

\begin{figure}[ht]
\includegraphics[width=\columnwidth]{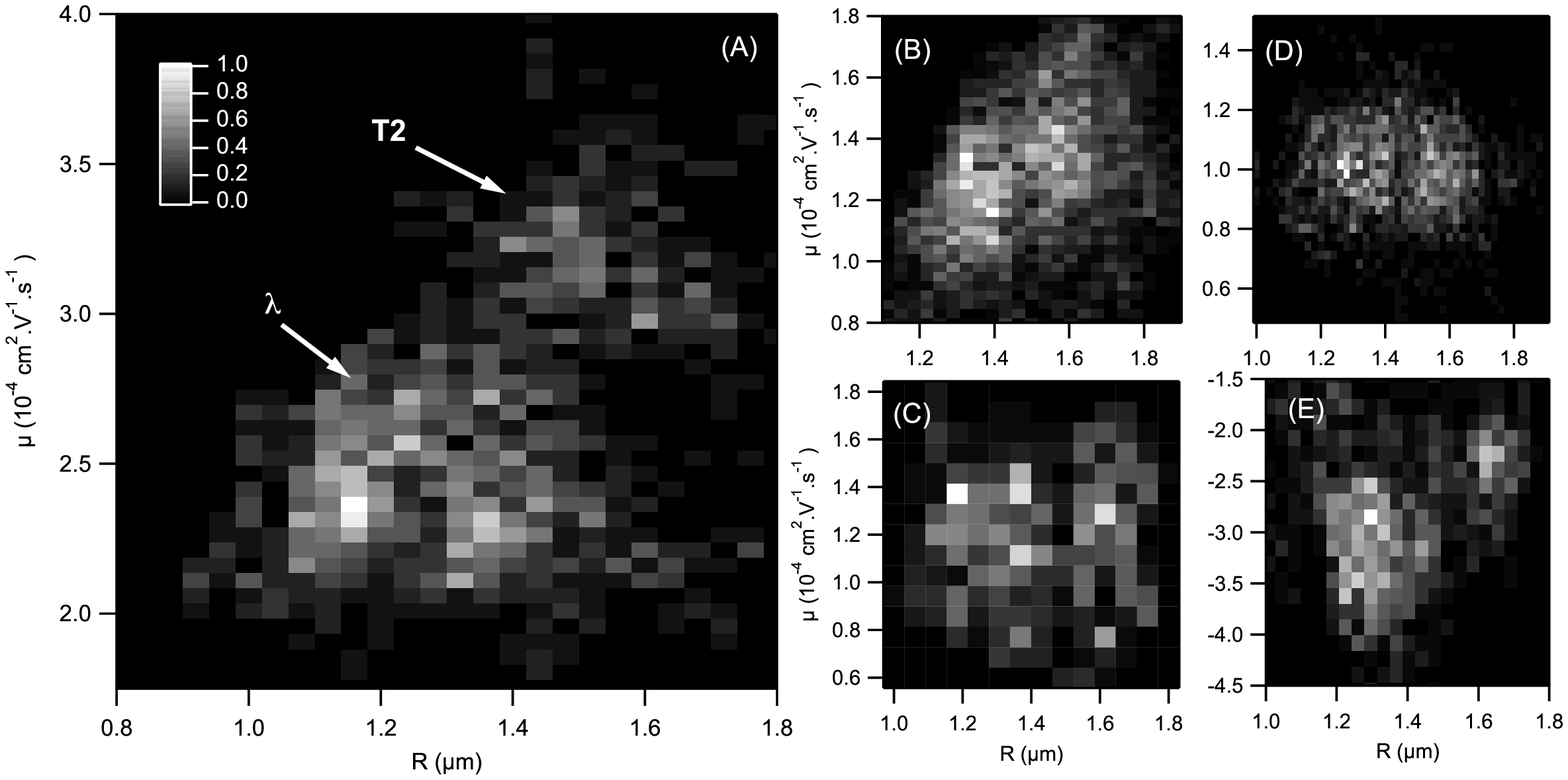}
\caption{
Normalized 2D histogram of the mobility of molecules {\em vs} their radius. Confinement thicknesses
are ($\mu$m): (A)  2.3, (B) 1.2, (C) 5, (D) 8 and (E) 3 with PDMA coating. Electric voltage is $60$ V. Bin
sizes are (units $\mu$m and $10^{-4}$ cm$^{2}\cdot$V$^{-1}\cdot$s$^{-1}$): (A) 0.04 and 0.05, (B) 0.04 and 0.03, (C) 0.06 and 0.08, (D) 0.02 and 0.03 and (E) 0.03 and 0.08. The mobilities are
defined as positive if the molecules move to the positive electrode. 
Notice the negative sign of the mobility when using PDMA coating which indicates that molecules go in
the opposite direction compared to the situation with EOF.
}
\label{f.3}
\end{figure}

In the second set of experiments, we have measured the mobilities in a 50-50 mixture of the original $T2$-DNA
and $\lambda$-DNA solutions. The weight concentration is thus still $1.2$ $\mu$g$\cdot$ml$^{-1}$, but since
$\lambda$-DNAs are 3 times shorter than $T2$-DNAs, the solution contains an average $T2$/$\lambda$ chain ratio
of 1:3. Fig. \ref{f.3} represents the 2D histogram of the velocity {\em vs} measured radius  for different
confinement thicknesses, showing two well-defined populations, with the expected 1:3 ratio. Note that the two
clusters are centered on about $1.3$~$\mu$m and about $1.6$~$\mu$m instead of the calculated $R_{g}$s
($0.56$~$\mu$m and $1.13$~$\mu$m). Deconvolution of the measured radii using the formalism of point
spread function leads to (resp.) 0.6 and $1.1$~$\mu$m, in excellent agreement with the calculated values.
The mean velocities of the two populations have been extracted from the centers of mass of each
population.

Mobility values are summed up in table \ref{table}. In the mixture, the overall behaviour is identical to the one
described for the first set of experiments (two regimes for the mobility and $T2$-DNA faster than
$\lambda$-DNA). Both $\lambda$-DNA and $T2$-DNA go faster when mixed. This observation is due to the procedure
of the determination of the mobilities in the mixture which leads to a larger uncertainty: we need to
distinguish two populations and then calculate the mobility of each population rather than measure the mean
mobility of the assembly of molecules seen on the images. We also observe that the smaller ($\lambda$) DNA chains move significantly faster in mixture than in pure solutions, maybe entrained by the larger ($T_2$) chains.

Finally, in order to prove that EOF indeed reverses the order of elution, we have coated the glass plates with
 poly(N,N,-dimethylacrylamide) (PDMA), a \emph{neutral} polymer, in order to eliminate the EOF. We let the
polymer adsorb on the glass surfaces by incubating the glass plates overnight in a PDMA solution (1\% w/w in
water)\cite{Barron}, then rinsed with distilled water. With the PDMA coating, we observe that the DNAs move
towards the anode (negative mobilities on fig. \ref{f.3}.E) which means that the EOF as been removed by the
coating. As expected, the absolute amplitudes of the two DNA mobilities are reversed: $\lambda$-DNA moves
faster than $T2$-DNA with mobilities of $+(3.1\pm 0.2) 10^{-4}$ cm$^{2}\cdot$V$^{-1}\cdot$s$^{-1}$ and $+(2.2\pm
0.2) 10^{-4}$ cm$^{2}\cdot$ V$^{-1}\cdot$s$^{-1}$  respectively.
\section{Discussion}

The experiments performed in capillaries and reported by Roeraade {\em et al.} \cite{Roeraade} and by Iki {\em et al.}
\cite{Iki} also shown a separation but with different features and interpretations as mentioned in the introduction.
But let us notice that our experimental conditions are significantly different:
\begin{itemize}
\item we have used a confinement between two plates and not in a capillary tube;
\item the diameters of the capillaries in  \cite{Roeraade} and \cite{Iki}  are always larger than the largest
chain fragments. Anyway, on decreasing the confinement (in our second regime), we never observe the behaviours
described by Roeraade and Iki;
\item they have used high electric fields (170 to 200 V$\cdot$ cm$^{-1}$)  which may lead to Joule heating,
resulting in radial convection in the capillary. Such an effect is not present in our experiments (maximum
field $\simeq 8$ V$\cdot$ cm$^{-1}$);
\item in the case of Roeraade's work, it was observed that a good separation is obtained only at low ionic strength (salt concentration close to 50 $\mu$M). 
\end{itemize}
All these differences explain that we observe original behaviours and we have thus to find different migration mechanisms. 
In the following, we present simple models combining the electric force applied on the DNA and the friction of
the molecule on the solvent and on the walls to explain the observed behaviours. 

In the second regime, the effective mobility $\mu$ remains constant ($= + (1\pm 0.3)  10^{-4}$ cm$^{2}\cdot$V$^{-1}\cdot$s$^{-1}$) and does not depend on the DNA length: this is characteristic of a bulk regime (no separation) with
an overall displacement of the solvent by the EOF. This yields:
\begin{equation}
\mu=\mu_{e}+\mu_{EOF}\label{MesMob}
\end{equation}
where $\mu_{e} < 0$ is the intrinsic mobility of the DNAs (in bulk solution and without EOF) and is mass
independent (free draining conditions). $\mu_{EOF}$ is related to the zeta potential $\zeta_S$ ($<0$) of the
glass surface by $\zeta_{S} = - (\eta\mu_{EOF})/(\epsilon_0\epsilon_r)$ with $\epsilon_0$ the vacuum
permittivity, $\epsilon_r$  and $\eta$ the dielectric constant and viscosity of the solution. Using  $\mu_{e}
= - 4\, 10^{-4}$ cm$^{2}\cdot$ V$^{-1}\cdot$ s$^{-1}$ \cite{Ekani,Stellw}(for a similar ionic strength),  we find
$\mu_{EOF}=+(5\pm 0.3)  10^{-4}$ cm$^{2}\cdot$V$^{-1}\cdot$s$^{-1}$, in agreement with the values reported in
\cite{Barron,Ekani,Stellw}($+ 5.6\; 10^{-4}$ cm$^{2}\cdot$V$^{-1}\cdot$s$^{-1}$) in similar conditions. This allows a
determination of the zeta potential of the glass plates: $\zeta_{S}  \simeq - 70$ mV (with $\eta =
10^{-3}$ N$\cdot$ m$^{-2}\cdot$s, $\epsilon_0= 8.8\;10^{-12}$ C$^2\cdot$ N$^{-1}\cdot$ m$^{-2}$ and $\epsilon_r
= 80$), in good agreement with reported values \cite{Zeta}.

On increasing the confinement (decreasing the gap $e$ between plates), we observe an important {\em increase}
of the mobility. Assuming that the EOF is constant (since $e$ is much larger than the Debye length
$\kappa^{-1} \simeq 2$ nm), this can be achieved solely by a decrease of the electric force acting on
DNAs. This may be realized likely through a local screening of the DNA charges within the proximal layers of
the glass plates but we have no model to discuss quantitatively this possibility.

In the first regime, the mobility increases with the gap thickness $e$ and decreases with the chain length
$N$, and our set-up separates DNAs of different length. We present a tentative theoretical model to describe the dependence of the effective mobility on the confinement thickness $e$ and the chain length (number of monomers $N$). 

Using the results of \cite{Long}, the effective mobility of the confined DNA is given by:
\begin{equation}
\mu=\frac{\mu_e+\mu_{EOF}}{1+\xi_1 /\xi}
\label{Theequation}
\end{equation}
$\xi_1\sim \eta N_1 a$ is the friction coefficient of the $N_1$ sliding monomers (of size $a$) on the cell walls and $\xi$ is the friction coefficient of the remaining $N-N_1$ monomers in the gap between the two walls.
In absence of specific adsorption, $N_1=N a/e$ \cite{DeGennes} and then $\xi_1\sim \eta N a^2/e$; inside the gap, the DNA chain behaves as an impermeable coil \cite{Long} and $\xi$ is proportional to $N^{1/2}$ (gaussian ideal chain) or to $N^{3/4} a^{5/4}e^{-1/4}$ (taking into account monomers excluded volume).  Eq. \ref{Theequation} then does predict an increase of the mobility with thickness $e$ but fails to describe the $N$ dependence.

Assuming $N_1 \sim N^{1/2}$, as suggested in \cite{Raphail} for unconfined DNA, we can satisfactorily describe the experimental
data if we consider that the chain inside the gap behaves as a free draining coil ({\em ie} $\xi \sim N$) but Eq. \ref{Theequation} then fails to describe the thickness dependence. We have experimentally addressed the possibility of strong DNA adsorption that could give special $N_1$
dependence on $N$. Using AFM (atomic force microscopy), we have determined that the maximum roughness of the
glass plates is about 2 nm: plate surfaces are smooth at the scale of the inter-plate gap size and DNA
separation cannot be attributed to chain trapping on surface defects. Furthermore we have not observed
\emph{J} or \emph{U} DNA shapes, characteristic of such trapping \cite{Austin}. Moreover, scanning the samples
in the $z$-direction with an optical microscope does not reveal any particular DNA adsorption on the glass
plates.

\section{Conclusion}

We have investigated the mobility of DNA chains confined between two solid walls.  The EOF reverses the order
of elution of the DNA chains. We have shown that steric confinement constitutes a simple and {\em fast} technique
for separation of large DNA, without gel or pulsed field, using weak electric fields. The interpretation of
the underlying mechanisms is still opened.

\section*{Acknowledgments} This work has been done while the authors were affiliated with the Institut Charles Sadron of
Strasbourg. The authors would like to thank Prof. Jean-Fran\c cois Joanny (Institut Curie) for fruitful discussions.

\newpage
\begin{table}[ht]
\label{table}
\begin{center}

\caption{Mobilities of $\lambda$- and $T2$-DNA for different confinements, measured in a pure solution and in
the mixture ($^{*}$interpolated).}
\label{table}
\vskip 1cm

\begin{tabular}{|c|c|c|c|}

\hline Species & Thickness & $\mu$    (pure)        & $\mu$ (mixture) \\ 
  & ($\mu$m) &  ($10^{-4}$ cm$^{2}\cdot$V$^{-1}\cdot$s$^{-1}$) & ($10^{-4}$ cm$^{2}\cdot$V$^{-1}\cdot
$s$^{-1}$)\\
\hline $\lambda$ DNA &  1.2    &  0.6                     &  1.3 $\pm$ 0.1 \\
\cline{2-4}
   & 2.3    &  1.7$^{*}$               & 2.4 $\pm$ 0.1 \\
\cline{2-4}
   & 5     &  0.8$^{*}$               &  1.2 $\pm$ 0.1 \\
\cline{2-4}
   & 8     & 1.1                      & 1.0 $\pm$ 0.1 \\
\hline\hline $T2$ DNA & 1.2       & 1.7                     & 1.4 $\pm$ 0.1\\
\cline{2-4}
 & 2.3       & 3.0$^{*}$               & 3.2 $\pm$ 0.1\\
\cline{2-4}
 & 5         & 1.1                     & 1.3 $\pm$ 0.1\\
\cline{2-4}
 & 8         & 1.1                     & 1.0 $\pm$ 0.1\\
\hline
\end{tabular}

\end{center}

\end{table}
\newpage
\indent

\indent

\begin{thebibliography}{0}

\bibitem{Slater}
{Slater G. W, Desruisseaux C., Hubert S. J., Mercier J.-F., Labrie J., Boileau J., Tessier F. \and P\'{e}pin M. P.}
{\em Electrophoresis} {\bf 21} (2000) {3873}

\bibitem{Raphail}
{Pernodet N., Samuilov V., Shin K., Sokolov J., Rafailovich M.H., Gersappe D. \and Chu B.}
{\em Phys. Rev. Lett.} {\bf 85} (2000) {5651}

\bibitem{Craig}
{Han J., Turner S.W. \and Craighead H.G.}
{\em Phys. Rev. Lett.} {\bf 83} (1999) {1688}

\bibitem{Roeraade}
{Roeraade J.  \and Stjernstr\"{o}m M.}
{International Patent WO/1997/26531 (available at www.wipo.int)} (1997)

\bibitem{Iki}
{Iki N., Kim Y. \and Yeung E.S.}
{\em Anal. Chem} {\bf 68} (1996) {4321}

\bibitem{Benoit}
{Benoit H. \and Doty P.}
{\em J. Phys. Chem}  {\bf 57} (1953) 958

\bibitem{hager}
{Hagermann P.J.}
{\em Annu. Rev. Biophysics Biophys. Chem.} {\bf 17} (1988) {265}

\bibitem{Grier}
{Crocker J.C. \and Grier D.G.}
{\em J. Colloid Interface Sci.} {\bf 179}(1996) {298}

\bibitem{Maier}
{Maier B. \and R\"{a}dler J.O.}
{\em Phys. Rev. Lett.} {\bf 82} (1999) {1911}

\bibitem{Barron}
{Barron A.E., Blanch H.W. \and Soane D.S.}
{\em Electrophoresis} {\bf 15} (1994) {597}

\bibitem{Ekani}
{Ekani Nkodo A. \and Tinland B.}
{\em Electrophoresis} {\bf 23} (2002) {2755}

\bibitem{Stellw}
{Stellwagen N.C., Gelfi C. \and Righetti P.G.}
{\em Biopolymers} {\bf 42} (1997) {687}

\bibitem{Zeta}
{Gu Y.G. \and Li D.Q.}
{\em J. Colloid Interface Sci.} {\bf 226} (2000) {328}

\bibitem{Long}
{Long D., Viovy J.-L. \and Ajdari A.}
{\em Phys. Rev. Lett.} {\bf 76} (1996) {3858}

\bibitem{DeGennes}
{de Gennes P.-G.}
{\em Scaling concepts in polymer physics},  {Cornell University Press} (1989)

\bibitem{Austin}
{Bakajin O.B., Duke T.A.J., Chou C.F., Chan  S.S., Austin R.H. \and Cox E.C.} 
{\em Phys. Rev. Lett.} {\bf 80} (1998) {2737}



\end{thebibliography}
\end{document}